\documentclass{article}

\usepackage{PRIMEarxiv}

\usepackage[utf8]{inputenc} 
\usepackage[T1]{fontenc}    
\usepackage{hyperref}       
\usepackage{url}            
\usepackage{booktabs}       
\usepackage{amsfonts}       
\usepackage{nicefrac}       
\usepackage{microtype}      
\usepackage{fancyhdr}       
\usepackage{graphicx}       
\graphicspath{{media/}}     
\usepackage{amsmath}
\usepackage{bm}

\title{Dynamic classifier auditing by unsupervised anomaly detection methods: an application in packaging industry predictive maintenance
}

\author{
  Fernando Mateo, Joan Vila-Francés, Emilio Soria-Olivas, Marcelino Mart{\'i}nez-Sober  \\
  \textbf{Juan G\'omez-Sanchis, Antonio José Serrano-L{\'o}pez}\\
  Intelligent Data Analysis Laboratory (IDAL) \\
  University of Valencia \\
  Valencia, Spain \\
  \texttt{\{fernando.mateo, joan.vila, emilio.soria, marcelino.martinez,} \\
  \texttt{juan.gomez-sanchis, antonio.j.serrano\}@uv.es} \\
}

\begin{document}
\maketitle

\begin{abstract}
Predictive maintenance in manufacturing industry applications is a challenging research field. Packaging machines are widely used in a large number of logistic companies' warehouses and must be working uninterruptedly. Traditionally, preventive maintenance strategies have been carried out to improve the performance of these machines. However, this kind of policies does not take into account the information provided by the sensors implemented in the machines. This paper presents an expert system for the automatic estimation of work orders to implement predictive maintenance policies for packaging machines. The key idea is that, from a set of alarms related to sensors implemented in the machine, the expert system should take a maintenance action while optimizing the response time. The work order estimator will act as a classifier, yielding a binary decision of whether a machine must undergo a maintenance action by a technician or not, followed by an unsupervised anomaly detection-based filtering stage to audit the classifier's output. The methods used for anomaly detection were: One-Class Support Vector Machine (OCSVM), Minimum Covariance Determinant (MCD) and a majority (hard) voting ensemble of them. All anomaly detection methods improve the performance of the baseline classifer but the best performance in terms of F1 score was obtained by the majority voting ensemble.  
\end{abstract}


\section{Introduction}

Production lines in large companies rely on equipment to work properly. A failure in a device or component may cause a stop in entire production line. The production stops are associated with huge costs, causing not only loss of production due to downtime, but also requiring efforts in identifying the cause of the failure and repairing it \cite{ayvaz2021predictive}. In the new industrial era, predictive maintenance has been adopted by many companies to estimate when maintenance should be performed to avoid unnecessary losses \cite{froger2016maintenance}.

In recent years, industry develops towards the fourth stage of industrialization \cite{stock2016opportunities} (the so-called Industry 4.0). The connectivity of all kinds of devices, particularly industrial devices, has experimented a substantial growth due, in part, to the improvement and cheapening of Ethernet-based buses. The current trend is that any industrial device, from a Programmable Logic Controller (PLC) to a complete manufacturing cell, is connected to a network. Usually, industrial devices send data related to the manufacturing process (performance, machine status, etc.). Thus, data analysis related to the improvement of industrial processes is indeed a reality. A proof of this is the wide and prolific state of the art in this field \cite{6748057, Liu201660}. The incorporation of advanced data-driven techniques and models allows to go beyond the simple visualization of process parameters. These techniques allow to obtain important information for the maintenance optimization and management of the industrial manufacturing cells.

\begin{figure}[!h]
\begin{center}
\mbox{\includegraphics[width=0.5\textwidth,keepaspectratio]{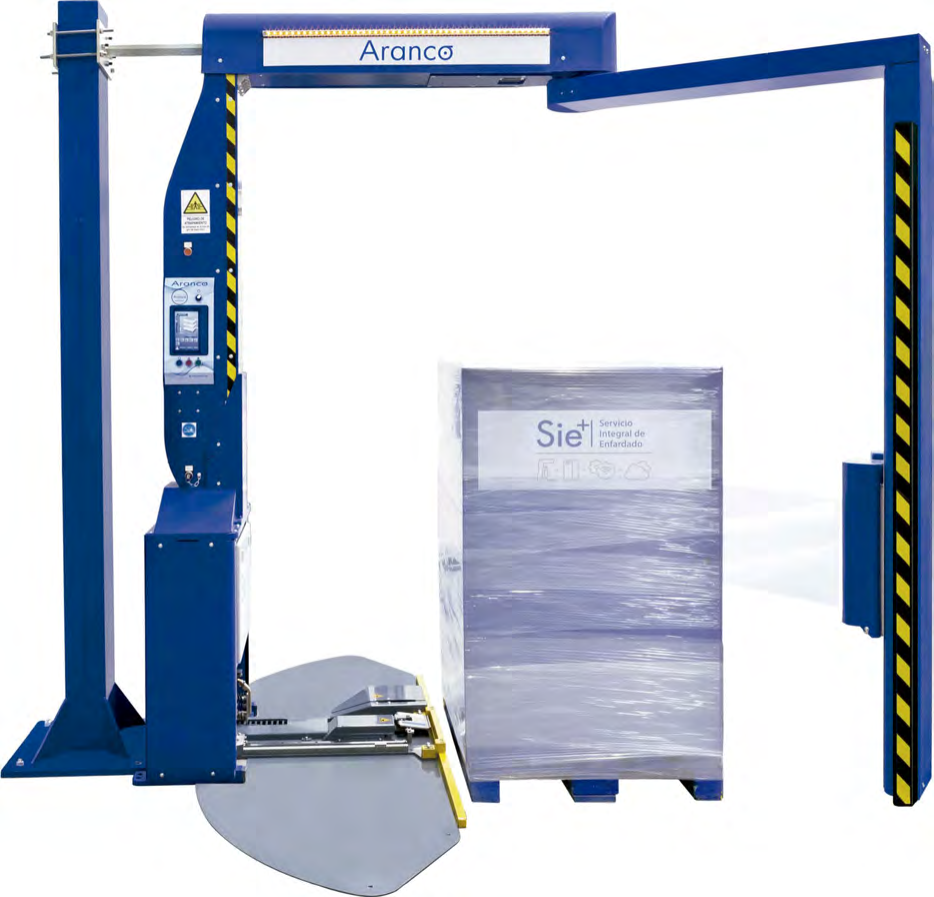}}
\caption[packagingmachine]{ \small Automatic packaging machine.}
\label{packagingmachine}
\end{center}
\end{figure}

This paper will address the problem of establishing predictive maintenance strategies in a group of packaging machines that work in logistic centers. Packaging machines are devices used for applying sheets of plastic film on transport pallets. These machines are designed to wrap the load without having to turn around a platform, in order to secure the load, protect it from dust or water, reduce its volume, strengthen the packaging, etc. An example of automatic fixed packaging machine is shown in Fig. \ref{packagingmachine}. All these packaging machines are connected to an Ethernet network and periodically report a set of alarms associated with different types of anomalous behaviours. Status variables of each machine are received by a central server and are stored in a database. Despite the availability of the status variables of each machine in real time, several challenges arise when planning the development of an expert system to decide when the company must send a technician to take maintenance actions. These challenges are related to some main issues: 1) as in almost every predictive maintenance problem there is a strong imbalance between the classes of the data set. This imbalance should be addressed before building a classifier, to avoid the undesirable situation where the classifier would tend to constantly predict the majority class, omitting all failure predictions \cite{lee2000noisy}; 2) the extraction of the relevant alarms and the construction of the data set that will be the input to the machine learning algorithms; 3) the definition of a strategy to determine the importance of past alarms in the classification system; 4) the frequency with which the models are updated and 5) the integration of the machine algorithms in the production lines of the manufacturing company.

The main contribution of this paper will be to introduce the framework and methodology to audit an expert system (a Random Forest-based classifier) for planning predictive maintenance actions (work orders) in the ERP (Enterprise Resource Planning) system of a packaging machine company, by appending an unsupervised anomaly detection stage at its output that acts as a supervisor. The anomaly detection stage only uses the information given by the classifier during the first 30 days of run time and is then applied seguentially to time windows of a determined duration to predict if the following day may be considered as an anomaly and would involve a maintenance action.  We concluded that this auditing stage is useful to improve the precision and recall of the baseline classifer, measured globally as the F1 score.

The rest of the paper is organized as follows: Section 2 reviews the existing literature in the field. Section 3 formulates the objective of the work and provides a graphical description of the system. Materials and methods are explained in Section 4, including the strategy to build the data set and the anomaly detection methods and libraries used to fine-tune the expert system. Section 5 shows the obtained classification results and Section 6 discusses and analyzes those results.

\section{Literature review}

Predictive maintenance has been gaining prominence in multidisciplinary research groups, proposing the creation and integration of lines of research related to data acquisition, infrastructure, storage, distribution, security, and intelligence \cite{zonta2020predictive}. The ability to predict the need for maintenance of assets at a specific future moment is one of the main challenges in Industry 4.0. New smart industrial facilities are focused on creating manufacturing intelligence from real-time data to support accurate and timely decision-making that can have a positive impact across the entire organization \cite{o2015big, muhuri2019industry}. In particular, the possibility of performing predictive maintenance contributes to enhancing machine downtime, costs, control, and quality of production \cite{froger2016maintenance, zonta2020predictive, ayvaz2021predictive}. 

Emerging technologies such as Internet of Things (IoT) and Cyber Physical Systems (CPS) are being embedded in physical processes to measure and monitor real-time data from across the factory, which will ultimately give rise to unprecedented levels of data production. This obliges to resort to Big Data technologies to manage these massive amounts of data \cite{lee2014recent, o2015big, rodriguez2016general}. 

Historically, preventive maintenance techniques have been typically used to minimize failures in manufacturing cells using maintenance strategies such as breakdown maintenance, preventive maintenance, and condition based maintenance including models and algorithms in manufacturing \cite{jardine2006review, lu2009predictive, Chouiki}. However, nowadays, other kinds of strategies are also considered. Examples of these techniques are cloud-based predictive maintenance \cite{Wang} or mobile agent technologies \cite{cucurull2009agent}. According to \cite{jardine2006review}, maintenance approaches able to monitor equipment conditions for diagnostic and prognostic purposes can be grouped into three main categories: statistical approaches, artificial intelligence approaches and model-based approaches. As model-based approaches need mechanistic knowledge and theory of the equipment to be monitored, and statistical approaches require mathematical background, artificial intelligence approaches, and machine learning techniques in particular, have been increasingly applied in predictive maintenance applications \cite{carvalho2019systematic, zhang2019data}.

Examples of these techniques are Multiple Classifiers \cite{susto2014machine}, Random Forests (RF) \cite{prytz2015predicting} or Support Vector Machines (SVM) \cite{Gryllias2012326, li2014improving,  Langone2015268} among many others \cite{carvalho2019systematic}. These strategies take into account issues like real-time alarm monitoring in the manufacturing cell and the general status of the machine in order to determine the best timing to take maintenance actions. The acquired data collects and saves the machine state by means of communication interfaces connected between the machine and the computing servers that host the maintenance algorithms. 

Anomaly detection refers to identifying data values that significantly deviate from typical behavior, which can be caused by various factors such as errors in the acquisition system or industrial equipment malfunction \cite{erhan2021smart}. Anomaly detection can be approached through centralized or distributed solutions and using statistical or machine learning (ML) methods. Statistical approaches are based on the distribution of variables, while ML provides techniques for handling high-dimensional data and identifying hidden relationships in complex environments \cite{nunes2023challenges}. While previous works typically rely on fully-labeled data, such scenarios are less common in practice because labels are particularly difficult (or impossible) to obtain during the training stage.  Furthermore, even when labeled data are available, there could be biases in the way samples are labeled, causing distribution differences. Such real-world data challenges limit the achievable accuracy of prior methods in detecting anomalies. Opposedly, unsupervised methods rely on the principles of self-supervised learning without labels. This framework aims at learning the steady state of a working machine and, from that knowledge, trying to predict anomalous behaviour. One example of algorithm that follows this principle is the One-Class Classifier \cite{fernandez2018learning}. This framework allows to work with very few training examples and to implement on-line anomaly detection solutions that prove to complement a classifier in the task of detecting different types of anomalies \cite{morselli2021anomaly}.

\section{Expert system description and objective of the work}

\begin{figure}[h]
\begin{center}
\mbox{\includegraphics[width=0.8\textwidth,keepaspectratio]{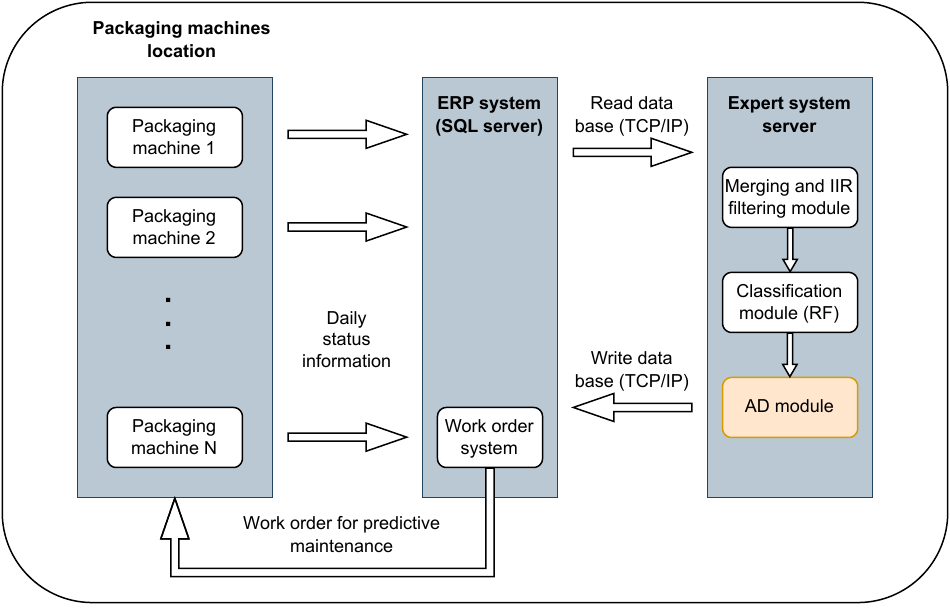}}
\caption{\small Predictive maintenance expert system. The main contribution of this work is the improvement of the existing classification module by means of unsupervised AD techniques (AD module).}
\label{expertsystem}
\end{center}
\end{figure}

Fig. \ref{expertsystem} shows the global structure of the proposed expert system. Packaging machine sensors report information about the status of the machine periodically to the ERP system of the company through a TCP/IP connection and an SQL database. This information is read by the expert system module and is preprocessed by the filtering module of the expert system in order to prepare the data to be used by the classification algorithms. Then, the classifier module writes the information related to the timing when a maintenance action has to be taken in the ERP SQL database. The objective of this work will be the development of an anomaly detection stage, following the existing pre-trained classifier, to optimize its precision and recall. The main novelty and contribution of this work is the development of dynamic unsupervised anomaly detection techniques which: a) avoid the tedious labeling process, and b) allow their swift deployment in a complex industrial maintenance and scheduling system for packaging machines without the need for a long training process. To achieve this goal, we developed and tested an on-line framework based on sliding windows in combination which several anomaly detection techniques.
 
\section{Methods}

\subsection{Data preprocessing}

The specification of the expert system imposed by the company was that it should provide a daily prediction to allow programming the daily schedule of working orders. Thus, the sampling period at which the patterns from each machine in the training set are acquired is one day. In order to build the data set, daily information about all the events from a machine (alarm occurrence, counters, working orders and tasks associated with these working orders) was collected, and then the information was organized in a vector containing all the information from that particular day. The alarm variables provide daily monitoring information that was used to train the classifier. An alarm event indicates some kind of malfunction in the machine. Table \ref{Tab_alarms} shows the posible causes of an alarm event.

\begin{table}[ht]
\centering
\begin{tabular}{ll}
\toprule
{\bf Alarm} & {\bf Description}\\
\midrule
A1 & Pulley failure \\
A2 & Arm engine failure \\
A3 & Maximun intensity in arm engine failure \\
A4 & Offset position failure\\
A5 & Communication failure \\
A6 & Minimun battery level failure \\
A7 & Maximun battery level failure  \\
A8 & Emergency button \\
A9 & Pulley failure \\
A10 & Carriage failure \\
A11 & Carriage engine failure \\
A12 & Vertical bar failure  \\
A13 & Horizontal bar failure \\
A14 & Maintenance  failure 1\\
A15 & Maintenance  failure 2 \\
A16 & Latch failure \\
A17 & Brake communication system failure\\
A18 & Plastic film broken  \\
A19 & Brake off\\
A20 & Excess strain on plastic film  \\
A200 & Communication error with remote board \\
A201 & Extended time without communication failure\\
\bottomrule
\end{tabular}
\caption{Set of alarms and their description.}
\label{Tab_alarms}
\end{table}  

Historical events since the previous working order are of paramount importance for the company. Therefore, a first order Infnite Impulse Response (IIR) filter was applied to the sensor signals from the machines to account for historical information. This filter weighs past information in a decreasing way, that is, the more recent the event, the more relevant is, according to Equation \ref{Eq_IIR}:

\begin{equation}
y(n) = \alpha\ y(n-1) + x(n)
\label{Eq_IIR}
\end{equation}

where $x(n)$ is the current input and $y(n-1)$ the previous output. The $\alpha$ parameter (0<$\alpha$<1) is used to weigh the past information. This filtering provides a decreasing sum of the frequency of ocurrence of each alarm from the date of the event to the last working order. The $\alpha$ parameter was empirically set to 0.63 after tests conducted by company experts. 

The target (binary) output is wether there was a maintenance action for that day. The classes in the data set were extremely imbalanced (1.3 \% positive (maintenance action required), 98.7 \% negative (maintenance not required). Therefore, the SMOTE method \cite{Chawla:2002:SSM:1622407.1622416} was employed to minimize the effect of class imbalance.

\subsection{Classifier}

The pre-trained classifier used by the packaging company was also developed by the authors of this article and consists of a random forest (RF) used to classify the binary variable that encodes whether a maintenance action took place on a particular date for a specific machine. RFs are a popular machine learning algorithm that can be used for both classification and regression tasks. They work by creating multiple decision trees, each trained on a random subset of the data and a random subset of the features. The final prediction is then made by averaging the predictions of all the individual trees \cite{breiman2001random}. Random forests are known for their high accuracy and robustness to overfitting. They have been successfully applied in various fields, such as finance, industry or medicine.

In particular, a single RF was trained using R caret package \cite{kuhn2005caret} to predict maintenance actions for all machines. The selected model is composed of 500 trees (\textit{ntree} parameter) and uses 17 ramdomly sampled predictors (\textit{mtry} parameter). The resulting Area Under the ROC is 0.848 on the test set (50 \% of the data). Theoretically, this result is satisfactory, but due to the significant imbalance between classes, a large number of false positives occur, resulting in a very low F1 Scores (approximately 0.2 in average).

\subsection{Anomaly detection methods}

Anomaly detection (AD) is the identification of observations that do not conform to an expected pattern or other items in a dataset. It is an important data mining tool for discovering induced errors, unexpected patterns, and more generally anomalies.

There are various ways of performing AD and the approach to take will depend on the nature of the data and the types of anomalies to detect. When labels are scarce or not available, unsupervised learning schemes may be considered. Basically, this model identifies outliers during the fitting process, where it learns a steady state and can then infer abnormal variations (or novelties) from it. It can be used when outliers are defined as points that exist in low-density regions of the dataset. Thus, any new observation which does not belong to high-density areas will be labeled as such automatically by the algorithm.

In this work two of the most successful AD metrods were compared, while also creating an ensemble of them. These are One-Class Support Vector Machine (OCSVM) and Minimum Covariance Determinant (MCD).

\subsubsection{OCSVM}

OCSVM is a variant of the traditional Support Vector Machine (SVM) that is trained on only one class of data, i.e., the normal data. The algorithm learns the boundaries of the normal data and identifies any data points that fall outside these boundaries as anomalous \cite{scholkopf1999support}. This algorithm can be used both in supervised or unsupervised manner.  OCSVM has been widely used in various fields such as finance, cybersecurity, medical diagnosis, and fault detection in industrial systems.

The OCSVM model used in this work is a wrapper of Python's scikit-learn one-class SVM class with more functionalities, that is included in the pyOD package \cite{zhao2019pyod}. In particular, we used an OCSVM with radial basis function (RBF) kernel. Given a dataset \( \{ \mathbf{x}_i \}_{i=1}^n \) where \( \mathbf{x}_i \in \mathbb{R}^d \), the objective of OCSVM with RBF kernel is to find the optimal hyperplane that separates the data from the origin, capturing the regions with the majority of the data points.

The decision function of the OCSVM with RBF kernel is given by:

\begin{equation}
f(\mathbf{x}) = sign\left(\sum_{i=1}^n \alpha_i \exp\left(-\frac{\|\mathbf{x} - \mathbf{x}_i\|^2}{2\sigma^2}\right) - \rho\right)
\end{equation}

\noindent where \( \sigma \) is the bandwidth parameter of the RBF kernel and $\rho$ represents the offset with respect to the origin. The primal optimization problem for OCSVM with RBF kernel can be formulated as:

\begin{equation}
\min_{\mathbf{w}, \rho, \boldsymbol{\xi}} \frac{1}{2} ||\mathbf{w}||^2 + \frac{1}{\nu n} \sum_{i=1}^n \xi_i - \rho
\label{Eq_primal_ocsvm}
\end{equation}

\noindent where $\nu$ controls the balance between outliers and support vectors and $\xi_i$ are the distances from each data point to the decision boundary. Equation \ref{Eq_primal_ocsvm} is subject to:

\begin{equation}
\begin{aligned}
& \mathbf{w}^T \boldsymbol{\phi}(\mathbf{x}_i) - \rho \geq 1 - \xi_i \\
& \xi_i \geq 0 \\
& i = 1, 2, ..., n
\end{aligned}
\end{equation}

By using a Lagrangian the problem can be solved in the dual space. The Lagrangian for this problem is:

\begin{equation}
\begin{aligned}
L(\mathbf{w}, \rho, \boldsymbol{\alpha}, \boldsymbol{\xi}) &= \frac{1}{2} ||\mathbf{w}||^2 + \frac{1}{\nu n} \sum_{i=1}^n \xi_i - \rho \\
&\quad - \sum_{i=1}^n \alpha_i (\mathbf{w}^T \boldsymbol{\phi}(\mathbf{x}_i) - \rho - 1 + \xi_i) - \sum_{i=1}^n \mu_i \xi_i
\end{aligned}
\end{equation}

And the dual problem is:

\begin{equation}
\max_{\boldsymbol{\alpha}} \min_{\mathbf{w}, \rho, \boldsymbol{\xi}} L(\mathbf{w}, \
max_{\rho, \boldsymbol{\xi}} L(\mathbf{w}, \rho, \boldsymbol{\alpha}, \boldsymbol{\xi})
\end{equation}

Subject to:

\begin{equation}
\begin{aligned}
& \alpha_i \geq 0 \\
& \mu_i \geq 0 \\
& i = 1, 2, ..., n
\end{aligned}
\end{equation}

After solving the dual problem, the optimal hyperplane is obtained by:
\begin{equation}
\begin{aligned}
\mathbf{w} &= \sum_{i=1}^n \alpha_i \boldsymbol{\phi}(\mathbf{x}_i) \\
\rho &= \text{median}\left(\{ \mathbf{w}^T \boldsymbol{\phi}(\mathbf{x}_i) \}_{i=1}^n\right)
\end{aligned}
\end{equation}

\noindent where \( \boldsymbol{\phi}(\mathbf{x}) \) denotes the feature mapping function.

Unlike linear or polynomial kernels, RBF is more complex and efficient at the same time that it can combine multiple polynomial kernels multiple times and using different degrees to project the non-linearly separable data into higher dimensional space so that it can be separable using a hyperplane.

\subsubsection{MCD}

The MCD estimator is another popular robust method for estimating multivariate location and scatter. It finds the best fitting elliptical distribution by minimizing the determinant of the covariance matrix subject to a constraint on the number of observations. First, it fits a minimum covariance determinant model and then computes the Mahalanobis distance as the outlier degree of the data. 

Supposing the data is stored in an $ n \times p$ matrix $\bm{X} = \{\bm{x}_1 \ldots \bm{x}_p\}^\prime$, the classical tolerance ellipse is defined as the set of $p$-dimensional points $\bm{x}$ whose Mahalanobis distance is defined in terms of statistical distance ($d$) as:

\begin{equation}
MD(\bm{x})=d(\bm{x}, \overline{\bm{x}}, Cov(\bm{X})) = \sqrt{(\bm{x}-\overline{\bm{x}})^\prime Cov(\bm{X})^{-1} (\bm{x}-\overline{\bm{x}})}
\end{equation}

\noindent equals the cutoff value $\sqrt{x^2_{p,0.975}}$ \footnote{We denote $x^2_{p,\alpha}$ as the $\alpha$ quantile of the $x^2_p$ distribution}. Here $\overline{\bm{x}}$ is the sample mean and $Cov(\bm{X})$  the sample covariance matrix. The Mahalanobis distance $MD(\bm{x}_i)$ should tell us how far away $\bm{x}_i$ is from the center of the data cloud, relative to its size and shape and, consequently, an outlier degree of the data point. On the other hand, the robust tolerance ellipse used by MCD is based on the robust distances:

\begin{equation}
RD(\bm{x})=d\left(\bm{x}, \hat{\bm{\mu}}_{\mathrm{MCD}}, \hat{\bm{\Sigma}}_{\mathrm{MCD}}\right),
\end{equation}

\noindent where $\hat{\bm{\mu}}_{\mathrm{MCD}}$ is the MCD estimate of location and $\hat{\bm{\Sigma}}_{\mathrm{MCD}}$  is the MCD covariance estimate.

The MCD estimator is resistant to outliers and can handle high-dimensional data. It  can also be applied both in supervised or unsupervised way. A study by \cite{rousseeuw1999fast} showed that the MCD estimator outperformed other robust estimators in terms of efficiency and breakdown point. Another study by \cite{croux1999influence} demonstrated the effectiveness of the MCD estimator in real-world applications.

The MCD model used in this work is a wrapper of Python's scikit-learn MinCovDet class function with more functionalities, that is included in the pyOD package \cite{zhao2019pyod}.

\subsubsection{AD ensembles}

Our initial findings when testing the aforementioned AD methods is that none of them obtains the best precision and recall in all scenarios. That is the reason why we propose to combine their outputs using a model ensemble. Several types of model ensembles exist in the literature, like voting, averaging, weighted averaging, etc. In this work we have selected a voting ensemble.

A voting ensemble works by combining the predictions from multiple models by returning the most voted result. It can be used for classification or regression. In the case of regression, this involves calculating the average of the predictions from the models. In the case of classification, the predictions for each label are summed and the label with the majority vote is predicted. In the binary case (normal or abnormal observation), a majority voting with two models involves that both models must agree that the observation is abnormal in order to tag it as an anomaly.

\subsection{Streaming framework}

Real-time AD is crucial in industrial environments as it helps in promptly recognizing and resolving errors before they cause major failure or even destruction of industrial equipment. It can also provide insight on the causes that lead to those failures based on the ongoing activities of a system in near real-time, enabling to make better decisions by detecting anomalies as they occur.

To enable a real-time implementation of the AD models in an industrial environment we propose a streaming framework based on sliding windows. For this purpose, this work profits from Python's pySAD library \cite{pysad}. This library is open-source and compatible with pyOD. The window scheme is based on the study by \cite{manzoor2018xstream} that addresses the outlier detection problem for feature-evolving streams using an algorithm called xStream. In our work we use a similar window scheme but using different detectors (OCSVM, MCD and voting ensemble).

Essentially, the AD models observe the probabilistic output from the classifier for a particular machine and for a determined period of time. This initial window allows to train the AD models in an unsupervised way. In this stage, the models extract information about the normal or abnormal status of the machine. This initial window size was set to 30, which corresponds to 30 daily data. During this training period, no output is given by the AD system.

\begin{figure}[h]
\begin{center}
\mbox{\includegraphics[width=0.9\textwidth,keepaspectratio]{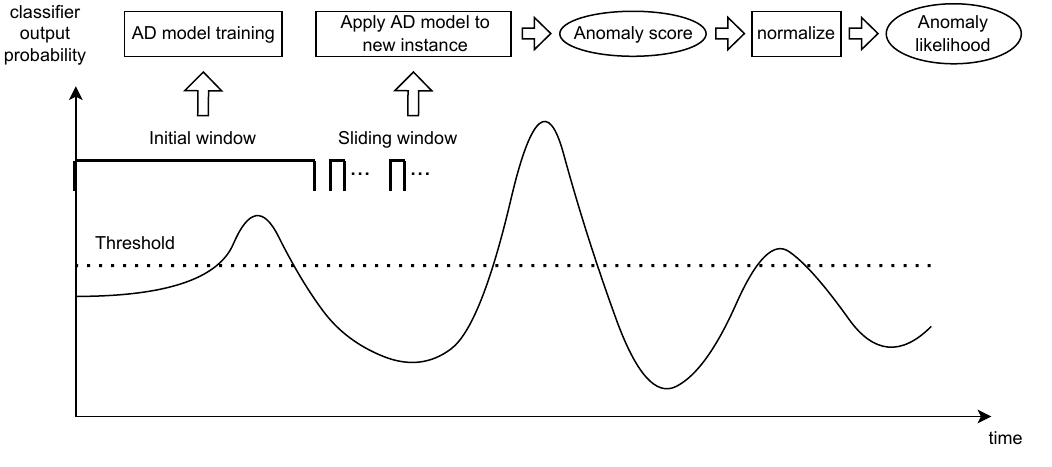}}
\caption{\small Anomaly detection streaming procedure for on-line auditing of the classifer.}
\label{streaming}
\end{center}
\end{figure}

\begin{figure}[h]
\begin{center}
\mbox{\includegraphics[width=0.4\textwidth,keepaspectratio]{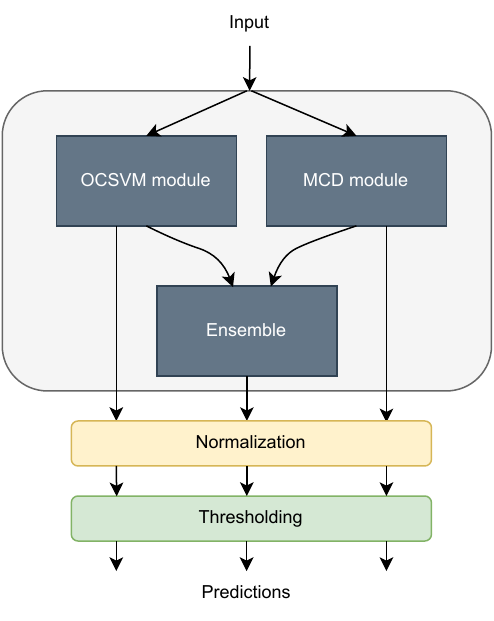}}
\caption{\small Proposed architecture for the AD module.}
\label{ADFlowchart}
\end{center}
\end{figure}

Once the initial window frame has passed, the probabilistic output from the classifier is fed sequentially to the AD models, which are fitted to the new observation and provide an anomaly score. This score is then normalized to the [0,1] range to provide a new probability value. The probability value is then compared with the same decision threshold used by the classifier (that was optimized during its training stage) to convert the returned probability to a binary value. The streaming dynamics of the proposed approach is shown in Fig. \ref{streaming} and the final architecture proposed for the AD module is summarized in Fig. \ref{ADFlowchart}.

\subsection{Performance metric}

The performance metric used to compare the existing classifier (baseline) with the proposed methods was the F1 score, which is a robust metric that combines both precision and recall performances:

\begin{equation}
F_1=2 \frac{precision \cdot recall}{precision + recall}
\end{equation}

All results are referred to the test set, i.e. after the initial training window, to evaluate the generalization capabilities of the AD methods.

\section{Results}

\begin{figure}[h]
\begin{center}
\mbox{\includegraphics[width=\textwidth,keepaspectratio]{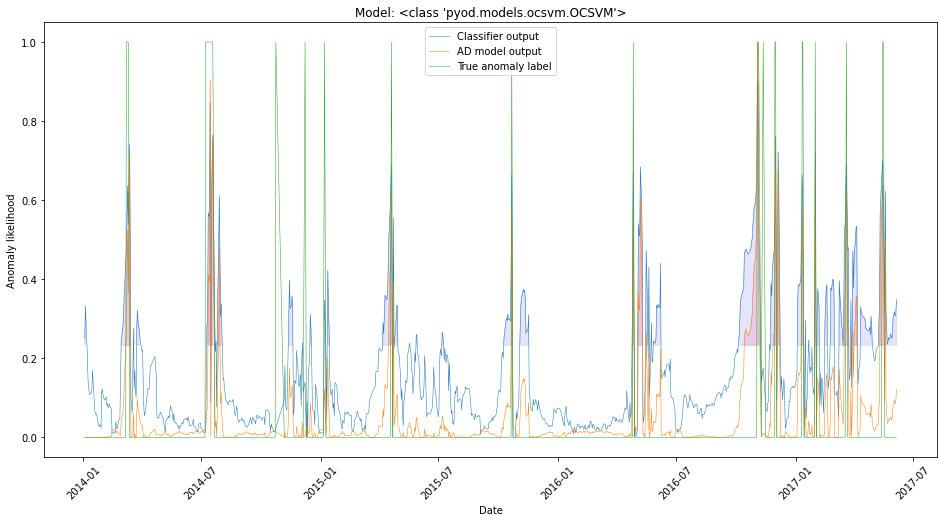}}
\caption{\small Auditing example for one of the machines using OCSVM. The output from the classifier and the AD model output are represented as different lines and the shaded part of each signal indicates if the threshold to call for a work order has been exceeded.}
\label{OCSVM_sample}
\end{center}
\end{figure}

We utilized machine records with the longest data history (more than 1000 days of run time) to test the proposed methodology. This approach ensured that the performance evaluation was realistic and reflective of long-term outcomes. This restriction reduced the number of machines to 23. The AD models were built and trained individually for each one of the machines. As an example, in Fig. \ref{OCSVM_sample} we represent the classifier output probability and the anomaly likelihood obtained OCSVM for one of the machines. The true labels are represented too in a differnt line. The threshold applied was obtained during the classifier's training to optimise the AUROC and determines the part of the signal where each method predicts a work order.

In this figure we observe that the proposed AD auditing stage starts producing a non-null output after the first 30 samples used to train the model. After that period, it filters the classifier output and makes it more selective when predicting a maintenance action. This behaviour is consistent with the rest of machines. The effect of the AD stage may me quantified by using a performance metric with respect to the true labels for each data point.

 The data used to calculate the score was in all cases the inferred data after the initial training period of 30 days. The obtained F1 scores for the three AD methods are compared in Table \ref{Performance}. This figure includes the baseline classifier F1 score to better assess the improvement offered by the AD methods.
In quantitative terms, the voting ensemble is the best method out of the tested, with an average test F1 score of 0.480 $\pm$ 0.17 across the 23 machines. OCSVM and MCD obtain an average F1 of 0.439 $\pm$ 0.17 and 0.44 $\pm$ 0.19, respectively. All methods improve significantly the baseline model, that produces a test F1 of 0.21 $\pm$ 0.10. The maximum absolute F1 score improvement was obtained for machine 25ARE2200:2AB-0118. Using the ensemble model, the F1 score was raised to 0.784 from a baseline of 0.277.

\begin{table}[ht]
\label{Performance}
    \centering
    \caption{Test set F1 score comparison between the three AD methods and the baseline classifier, for 23 packaging machines. The best results for each machine have been highlighted in bold. The maximum relative percentage change is indicated in the rightmost column. } 
    \vspace{0.5cm}
    \begin{tabular}{lrrrrr}
    \toprule
        \textbf{Machine} & \textbf{Baseline} & \textbf{OCSVM} & \textbf{MCD} & \textbf{Ensemble} & \textbf{max. \% change} \\ \midrule
        25ARE2200:2AB-0118 & 0.277 & 0.600 & \textbf{0.784} & \textbf{0.784} & +182.7\%  \\ 
        25ARE2200:2AB-0140 & 0.244 & 0.615 & \textbf{0.650} & \textbf{0.650} & +166.6\% \\ 
        25ARE22V2:2BC-0248 & 0.022 & 0.049 & 0.133 & \textbf{0.222} & \textbf{+933.3\%}  \\ 
        25ARE22V2:2BC-0264 & 0.235 & \textbf{0.620} & 0.533 & 0.612 & +164.3\% \\ 
        25ARE22V2:2BC-0268 & 0.321 & 0.545 & \textbf{0.596} & \textbf{0.596} & +85.9\%  \\ 
        25ARF2200:101-0020 & 0.167 & \textbf{0.408} & 0.208 & \textbf{0.408} & +144.9\% \\ 
        25ARF2200:101-0027 & 0.047 & 0.240 & 0.261 & \textbf{0.273} & +481.8\% \\ 
        25ARF2200:101-0035 & 0.288 & \textbf{0.756} & \textbf{0.756} & \textbf{0.756} & +162.6\% \\ 
        25ARF22V2:1AA-0042 & 0.119 & 0.267 & \textbf{0.286} & \textbf{0.286} & +140.0\% \\ 
        25ARF22V2:1AA-0083 & 0.129 & 0.250 & \textbf{0.333} & \textbf{0.333} & + 158.3\% \\ 
        25ARF22V2:1AA-0094 & 0.232 & 0.447 & \textbf{0.595} & \textbf{0.595} & +156.5\%  \\ 
        25ARF22V2:1AA-0098 & 0.125 & 0.571 & 0.171 & \textbf{0.600} & +380\%  \\ 
        25ARF22V2:1AA-0099 & 0.278 & 0.425 & \textbf{0.436} & \textbf{0.436} & +56.5\%  \\ 
        25ARF22V2:1AA-0108 & 0.392 & \textbf{0.419} & 0.372 & 0.381 & +6.9\% \\ 
        25ARF22V2:1AA-0109 & 0.238 & 0.308 & \textbf{0.370} & \textbf{0.370} & +55.6\%  \\ 
        25ARF22V2:1AA-0110 & 0.212 & 0.310 & \textbf{0.333} & \textbf{0.333} &  +56.9\% \\ 
        25ARF22V2:1AA-0113 & 0.140 & 0.421 & 0.356 & \textbf{0.516} & +267.7\% \\ 
        25ARF22V2:1AA-0173 & 0.195 & 0.438 & \textbf{0.524} & \textbf{0.524} & +167.9\% \\ 
        25ARF22V2:1AA-0176 & 0.145 & \textbf{0.367} & 0.200 & 0.200 & +153.2\% \\ 
        25ARF22V3:1BB-0184 & 0.118 & \textbf{0.400} & \textbf{0.400} & \textbf{0.400} & +240\% \\ 
        25ARF22V3:1BB-0188 & 0.127 & 0.364 & \textbf{0.381} & \textbf{0.381} & +200\% \\ 
        25ARF22V3:1BB-0189 & 0.341 & 0.604 & \textbf{0.672} & \textbf{0.672} & +97.0\% \\ 
        25ARF22V4:1CC-0308 & 0.357 & 0.676 & \textbf{0.701} & \textbf{0.701} & +96.5\% \\ 
        \midrule
        Average & 0.206 & 0.439 & 0.437 & \textbf{0.480} & +198.0\% \\ 
        \bottomrule
    \end{tabular}
\end{table}

\begin{figure}[ht]
\begin{center}
\mbox{\includegraphics[width=0.8\textwidth,keepaspectratio]{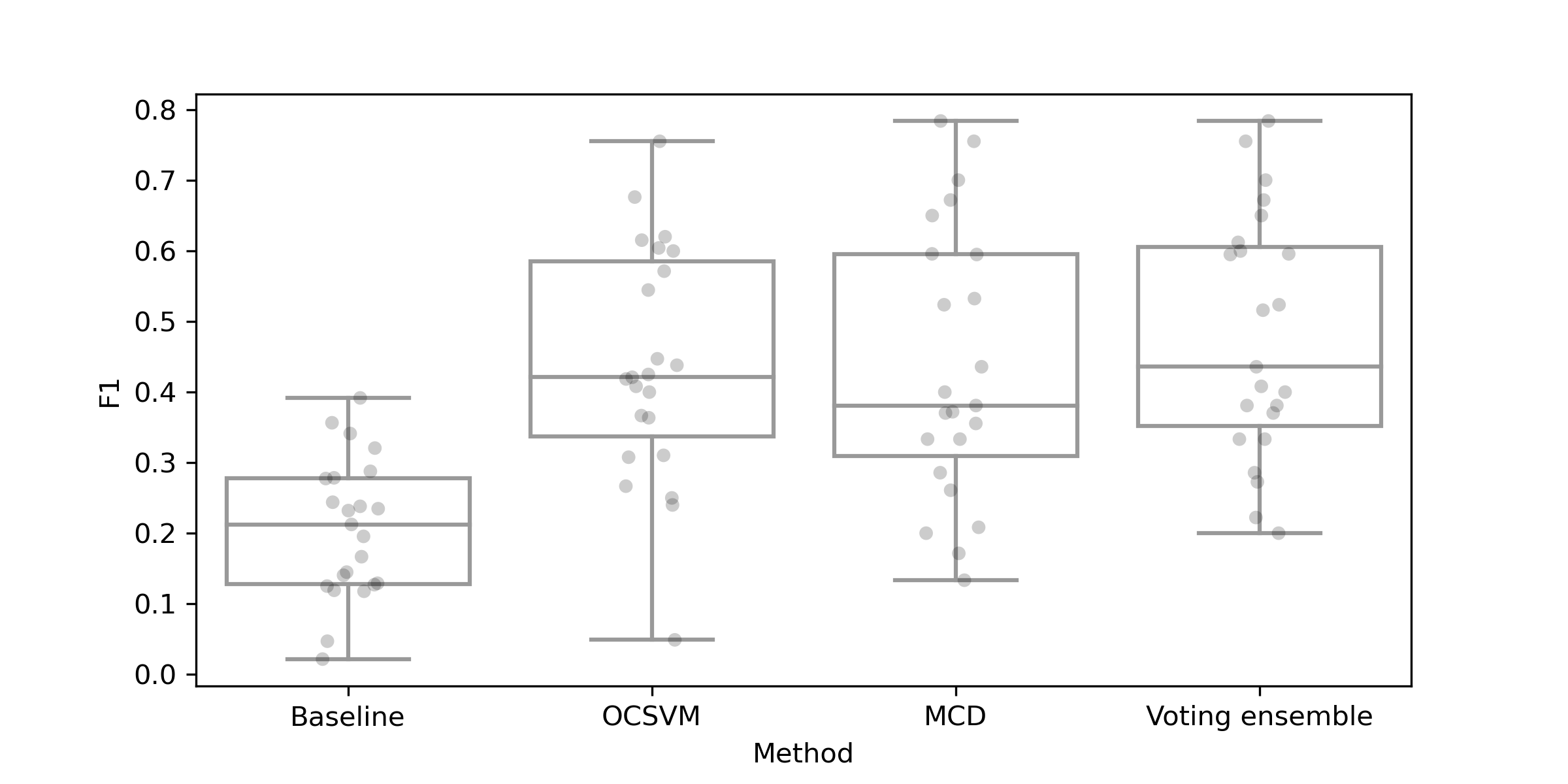}}
\caption{\small Box plots illustrating the differences between the F1 scores distributions obtained by the compared methods.}
\label{Boxplots}
\end{center}
\end{figure}

A statistical analysis was conducted to determine statistical significances about the differences between the three proposed methods and the baseline. Fig. \ref{Boxplots} shows the differences between the F1 distributions for each method. From the distributions, it is immediate to tell that the three proposed methods are statistically different from the baseline. The voting ensemble is the superior method in terms of median followed by OCSVM.  
A Shapiro-Wilk test demonstrates the normality of each group (p $>$ 0.1) and the one-way ANOVA test indicates significant differences between the means (p $<$ 0.001).

\section{Discussion}

One of the key challenges in predictive maintenance for industry is the accurate identification of anomalous events or failures to prevent costly downtime. In this context, OCSVM and MCD methods have emerged as effective techniques to audit and correct the output of classifiers. OCSVM is a powerful algorithm that is widely used for anomaly detection tasks. It constructs a hyperplane that separates normal data points from outliers in a high-dimensional feature space, leveraging only the information from the normal class. The method has shown promising results in various industrial applications, including fault detection and diagnosis. On the other hand, MCD is a robust estimator of multivariate location and scatter, which makes it suitable for detecting outliers in datasets with high-dimensional features. It fits a Gaussian distribution to the data, and outliers are identified as observations with low probabilities under the fitted model. Numerous studies have demonstrated the effectiveness of OCSVM and MCD methods in auditing and correcting classifier outputs for predictive maintenance tasks, improving the overall accuracy and reliability of anomaly detection. These methods provide a valuable means of refining the results of classifiers, enhancing the decision-making process, and facilitating more efficient maintenance strategies in industrial settings. However, according to our findings, none of the aforementioned methods has an absolute superiority over the other, so they need to be evaluated separately and their performance needs to be compared. 

We also proposed an ensemble method that tags a sample as an outlier only if both methods are in agreement (voting). Ensemble methods have gained significant popularity in various fields due to their ability to improve the robustness and overall performance of predictive models compared to individual methods. One key advantage of ensemble methods is their capacity to reduce the risk of overfitting by combining multiple models. By aggregating the predictions of these models, ensemble methods can effectively smooth out errors and biases that may be present in individual models, leading to more accurate and robust predictions. Moreover, ensemble methods can handle complex relationships and capture non-linear patterns by combining different modeling techniques or incorporating diverse feature representations. This versatility allows ensembles to adapt to a wide range of data distributions and increase the generalization capability of the model. In addition, ensemble methods can provide a better exploration of the feature space and overcome the limitations of individual models, thus enhancing the model's ability to capture important and relevant patterns.

On paper, the high AUC obtained on the test set by the original classifier previously developed by our group is satisfactory. However, when the client company deployed it to predict work orders in the ERP system, due to the class imbalance, no decision threshold on the signal captured from each machine is optimal: it either prioritizes false positive detection or true negative detection, which results in a low F1 score. The main advantage of adding the proposed AD stage is the significant increase in the F1 score without modifying the input information to the model. Since the model is in production, this aspect is of paramount importance.

From the results of our experiments, we observe that all AD techniques contribute to improve the classifier performance. In particular, the ensemble method has obtained the best results in terms of F1 score, followed by MCD and OCSVM, in this order. Out of the 23 machines, the ensemble method obtains the best performance in 20 cases, although it ties with MCD in 14 of those cases. In 4 cases it improves the performance of MCD or OCSVM alone. With regard to OCSVM, despite not being the best method in many cases, it achieves a similar average F1 score to that of MCD, and has shown potential to improve beyond the ensemble performance in 3 of the cases.

The percentage of F1 improvement with respect to the baseline model is leaded by the ensemble method, with an average improvement of 192.5 \%, followed by MCD with an improvement of 146.7 \% and OCSVM with 140.1 \%.
The absolute maximum F1 improvement was achieved by the ensemble for machine 25ARE22V2:2BC-0248 (933.3 \%) while the absolute minimum was obtained by MCD for machine 25ARF22V2:1AA-0108	(-5.0 \%).  However, this decrease in F1 with respect to the baseline only happens with one machine, so it is can be considered a very unlikely scenario.

\section{Conclusions}

This paper proposes a framework to audit the classifier implemented in an expert system used for planning predictive maintenance policies in the ERP (Enterprise Resource Planning) system of a packaging machine company.

The study was conducted on 23 time series with more than 1,000 data points, corresponding to daily predictions of the degree of likelihood of a work order. An unsupervised AD module was cascaded at the pretrained classifier's output so that, for every new sample, it produces three outputs (based on OCSVM, MCD and an ensemble of them). These predictions were used to audit and correct the classifier's output. This new score is normalized and thresholded to determine if the sample actually corresponds to an anomaly.

By examining the results, we observe that all AD techniques contribute to improve the classifier performance. More specifically, the ensemble method has obtained the best results in terms of F1 score with an average F1 increase of 192.5 \% on the test set, although the individual AD methods also produced an average increase above 140 \%.

The main contribution of the proposed methods reside in their simplicity, speed and ability to audit the classifier in real time, producing an output every time a new sample arrives. This fact combined with the significant increase in the precision and recall for an already optimised classifier yields an extremely interesting improvement for this particular system and could be easily transferred to other similar predictive maintenance settings. 

\section*{Acknowledgments}
This research is supported by the grant PID2021-127946OB-I00 funded by MCIN/AEI/ 10.13039/501100011033 by “ERDF A way of making Europe”. 

\bibliographystyle{unsrt}  
\bibliography{references2}

\end{document}